\documentclass[aps,nofootinbib,preprint]{revtex4}

\usepackage{graphicx}

\usepackage{amsmath}
\usepackage{amsfonts}
\usepackage{amssymb}
\usepackage{hyperref}
\usepackage{color}

\newcommand{\be}{\begin{equation}}
\newcommand{\ee}{\end{equation}}

\def\beq{\begin{equation}}
\def\eeq{\end{equation}}
\def\bea{\begin{eqnarray}}
\def\eea{\end{eqnarray}}

\def\slashchar#1{\setbox0=\hbox{$#1$}           
   \dimen0=\wd0                                 
   \setbox1=\hbox{/} \dimen1=\wd1               
   \ifdim\dimen0>\dimen1                        
      \rlap{\hbox to \dimen0{\hfil/\hfil}}      
      #1                                        
   \else                                        
      \rlap{\hbox to \dimen1{\hfil$#1$\hfil}}   
      /                                         
   \fi}

\begin{document}

\vspace*{1cm}
\title{On the production of a composite Higgs boson}

\author{\vspace{0.5cm}Ian Low$\,^{a,b}$ and Alessandro Vichi$\,^c$}
\affiliation{\vspace{0.5cm}\mbox{$^{a}$High Energy Physics Division, Argonne National Laboratory,
Argonne, IL 60439, USA}\\ 
\mbox{$^{b}$Department of Physics and Astronomy, Northwestern University,
Evanston, IL 60208, USA}\\
\mbox{$^{c}$Institut de Th\'eorie des Ph\'enom\`enes Physiques, EPFL, CH-1015 Lausanne, Switzerland}\vspace{0.8cm}
}

\begin{abstract}
\vspace*{0.2cm} 
We present a model-independent prescription for computing the gluon fusion production rate of a composite Higgs boson, which arises as a pseudo-Nambu-Goldstone boson, using effective lagrangians. The calculation incorporates three different effects due to the composite nature of the Higgs, some of which were neglected previously.  We apply the prescription to models with and without the collective breaking mechanism.  In sharp contrast with the case of a fundamental Higgs scalar, the rate only depends on the decay constant $f$ and is not sensitive to masses of new particles. After including electroweak constraints, there is a substantial reduction in the rate, in the range of 10 -- 30 \% or greater.

\end{abstract}


\maketitle


\section{Introduction}

The Higgs boson is the last missing piece in the standard model (SM) of particle physics, whose existence is all but assured by the precision electroweak data \cite{:2005ema}. The hunt for the Higgs is currently under way at the Tevatron and the Large Hadron Collider (LHC). Due to the unique role of the Higgs in the electroweak symmetry breaking, its discovery will have implications far beyond the mere observation of yet another new particle. In fact,  there is high hope that the Higgs will be the passage to a complete understanding of  physics at the TeV scale and beyond.

Measurements of the Higgs mass, as well as various production cross-sections and decay branching ratios, will provide important constraints in formulating a theory of electroweak symmetry breaking. In hadron colliders such as the Tevatron and the LHC, the most important production mechanism is the gluon fusion channel, which relies on the effective coupling of the Higgs to two gluons. This coupling, as it turns out, only arises at the one-loop level in minimally coupled theories, and receives a contribution from colored particles with a significant coupling to the Higgs. In the SM there is only one such particle: the top quark, which indeed gives the dominant contribution  in the SM. However, in many theories beyond the SM there are additional colored particles which could in principle have substantial impact on the Higgs production rate. Indeed, in supersymmetric theories and universal extra-dimensional models the gluon fusion rate is very sensitive to the masses of the top squark \cite{Dermisek:2007fi} and the first Kaluza-Klein top quark \cite{Petriello:2002uu}, respectively. Historically this is the effect that is studied most extensively. Recently there are efforts to systematically compute higher order QCD effects of new colored particles, according to their quantum numbers, in the gluon fusion channel \cite{Boughezal:2010ry}. For a recent update on the status of theoretical predictions on the Higgs production within the SM, see Ref.~\cite{Boughezal:2009fw}.

\begin{figure}[t]
\includegraphics[scale=0.65]{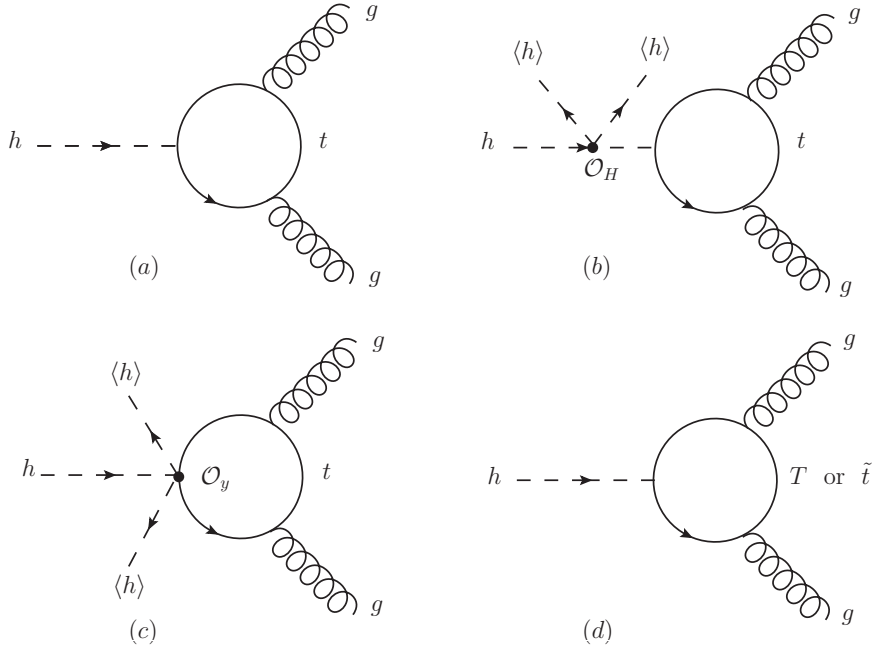} 
\caption{\em \label{fig:fig1} Diagrams $(a)$ and $(d)$ are the contributions to ${\cal O}_g$ from the SM top and a heavy fermions $T$ or scalar $\tilde{t}$ . Diagrams $(b)$ and $(c)$ summarize the effects of ${\cal O}_H$ and ${\cal O}_y$.  }
\end{figure}

In addition to new colored particles, the Higgs coupling to two gluons could be modified by other new physics effects in a less transparent fashion, which nevertheless could turn out to be equally, and in some cases more, important than effects of new colored particles. To understand these other effects, it is most instructive to use the method of effective lagrangians \cite{Weinberg:1978kz} by assuming all new physics effects, including that of new colored particles, could be encoded in a series of higher dimensional operators made out of the SM fields. Then the gluon fusion production of the Higgs is controlled by the following three operators \cite{Giudice:2007fh}:
\be
\label{eq:threeope}
{\cal O}_H = \partial^\mu (H^\dagger H)\partial_\mu(H^\dagger H) \ , \quad {\cal O}_y = H^\dagger H \bar{f}_L H f_R  \ , \quad {\cal O}_g =  H^\dagger H G_{\mu\nu} G^{\mu\nu} \ .
\ee
The operator receiving the most attention in the literature is ${\cal O}_g$, which results from integrating out the colored particles, including the SM top quark, in the loop. Although formally the procedure requires $m_h < 2m_t$, in practice it is found that the approximation works well for the Higgs mass up to 1 TeV \cite{Spira:1995rr}. Therefore, in this framework the production rate has no dependence on the Higgs mass at all. The operator ${\cal O}_H$ enters  because it contributes to the Higgs kinetic term, $\partial_\mu h\partial^\mu h$, after electroweak symmetry breaking, while ${\cal O}_y$ modifies the Higgs coupling to two fermions such as the top quarks. Diagrammatically, the effects of the three operators are summarized in Fig.~\ref{fig:fig1}, where diagrams (a) and (d) correspond to contributions to ${\cal O}_g$ upon integrating out the SM top and a new colored particle, respectively, while (b) and (c) demonstrate how ${\cal O}_H$ and ${\cal O}_y$ could alter the Higgs coupling to two gluons. In order to compute the Higgs production rate in the gluon fusion channel, all three operators must be taken into account.

In this work we present a model-independent prescription for computing the coefficients of the three operators for a composite Higgs boson, which arises as a pseudo-Nambu-Goldstone boson (PNGB). This class of models, along with weak-scale supersymmetry, are the two most attractive ideas for addressing the quantum stability in the Higgs boson mass. The proposal that the Higgs might be a PNGB, \`{a} la the pion of low-energy QCD, was first considered in Refs.~\cite{Kaplan:1983fs,Kaplan:1983sm} and later revitalized by the advance of little Higgs theories \cite{Arkani-Hamed:2001nc, Arkani-Hamed:2002qx,Arkani-Hamed:2002qy} and holographic Higgs models \cite{Contino:2003ve, Agashe:2004rs}.  The importance of the gluon fusion production rate in understanding the composite nature, or the lack thereof, of the Higgs was emphasized in Ref.~\cite{Low:2009di}, where it was shown that this production rate is reduced from the SM expectation in a generic composite Higgs model and enhanced in models where the Higgs mass is fine-tuned.\footnote{For examples of enhancements in the Higgs production in models where the Higgs mass is fine-tuned, see Refs.~\cite{Petriello:2002uu, Kribs:2007nz,Boughezal:2010ry}.} 

Although there have been several studies on the production rate of a composite Higgs \cite{Han:2003gf, Chen:2006cs, Djouadi:2007fm, Falkowski:2007hz, Maru:2010ic, Espinosa:2010vn, Casagrande:2010si,Azatov:2010pf}, these studies employed different approaches which may depend on the specific realizations of the pseudo-Goldstone nature of the Higgs boson. Our goal here is to demonstrate the conceptual generality and simplicity of the effective lagrangian approach, which does not depend on the particular mechanism stabilizing the mass of the Higgs boson. In some cases we improve on previous calculations ignoring part of the effects we consider here and compute the production rate properly, while in some other cases we re-derive the same results and justify {\em a posteriori} the approximations made in previous works. In addition, we also provide results on models in which the Higgs production rate have not been computed previously.

This paper is organized as follows: in Section \ref{sect:powercount} we review the effective lagrangian approach to compute the Higgs production rate, followed by a prescription to compute $c_H$, $c_y$, and $c_g$ in Section \ref{sect:prod}. After that we apply the prescription to several models with and without collective breaking mechanism in Section \ref{sect:models}. Then  in Section \ref{sect:discuss} we close out with the summary and discussions.

\section{Effective lagrangian in the Higgs production}
\label{sect:powercount}

We briefly overview the effective lagrangian approach \cite{Weinberg:1978kz} used to compute the Higgs production rate in this study. The approach is quite general and applies regardless of whether the Higgs is composite or not, as long as the new particles interacting with the Higgs are slightly heavier than the weak scale and can be integrated out when studying properties of the Higgs boson. Although we only consider models with one Higgs doublet, the assumption will hold in multi-Higgs-doublet models as long as one Higgs doublet is much lighter than all the other doublets, such as in the decoupling limit \cite{Gunion:2002zf} of the two-Higgs-doublet model.

Defining the Higgs doublet $H$ as
\be
\label{eq:hdoulbet}
H = \frac1{\sqrt{2}} \left(
\begin{array}{c}
h^+ \\
h^0
\end{array} \right) =
 \frac1{\sqrt{2}} \left(
\begin{array}{c}
h^1+ i h^2 \\
h^3+ih^4
\end{array} \right) \ ,
\ee
the effective lagrangian relevant for the gluon fusion production of the Higgs is, up to dimension-six,
\begin{eqnarray}
\label{L_SILH}
{\cal L}_{\rm eff}&=& (D_\mu H)^\dagger D^\mu H + \left(y_f \bar{f}_L H f_R + {\rm h.c.} \right) \nonumber \\
&+& \frac{c_H}{2f^2} \partial_\mu(H^\dagger H) \partial^\mu(H^\dagger H) + \left(\frac{c_y y_f}{f^2} H^\dagger H \bar{f}_L H f_R + {\rm h.c.}\right)    + \frac{c_g \alpha_s}{4\pi} \frac{y_t^2}{m_\rho^2}  H^\dagger H G_{\mu\nu} G^{\mu\nu} ,
\end{eqnarray}
where $f$ and $m_\rho$ are two generic mass scales related to the scale of new physics, and $y_f$ is the SM Yukawa coupling. In particular, $y_t$ is the SM top Yukawa coupling.  The operator with coefficient $c_H$ gives a contribution to the (neutral) Higgs kinetic energy, after the electroweak symmetry breaking, which requires an overall re-scaling of the Higgs to go back to the canonically normalized kinetic term.\footnote{\label{cusfoot}We neglect a custodially violating operator, $|H^\dagger \tensor{D}_\mu H|^2$, which does not contribute to the neutral Higgs kinetic term upon electroweak symmetry breaking but nonetheless shifts the $Z$ mass. This operator is severely constrained by electroweak $\rho$ parameter.} The term with $c_y$ modifies the Higgs coupling to the SM fermion, in particular the top quark. The last operator in Eq.~(\ref{L_SILH}) represents contributions from new heavy colored particles to the Higgs-gluon-gluon coupling, which is only induced at the one-loop level and hence the $\alpha_s$ suppression in its coefficient. Given a specific model, the dimensionless coefficients $c_H$,  $c_y$, and $c_g$ can be readily computed.

In Eq.~(\ref{L_SILH}) we have ignored an operator,
\be
 \frac{c_r}{2f^2} {\cal O}_r \equiv  \frac{c_r}{2f^2} H^\dagger H (D_\mu H)^\dagger D^\mu H \ ,
 \ee
which is eliminated by a field re-definition. This is the operator basis adopted in the SILH lagrangian \cite{Giudice:2007fh}. Parameters in a general operator basis, where $c_r\neq 0$, are related to the SILH basis  by 
\begin{equation}
\left. \phantom{C^{A^a}_{B_B}}c_H\right\vert_{SILH}=c_H-\frac{c_r}{2}, \qquad\qquad \left. \phantom{C^{A^a}_{B_B}}c_y\right\vert_{SILH}=c_y+\frac{c_r}{4}.
\end{equation}
It goes without saying that  physical amplitudes only depend on the reparameterization-invariant combination of parameters. One example is the combination $c_H+2c_y$,\begin{equation}
\left. \phantom{C^{A^a}_{B_B}}(c_H+2c_y)\right\vert_{SILH}=c_H+2c_y\,.
\end{equation}
which controls the on-shell Higgs coupling to the fermion \cite{Giudice:2007fh}.

Including the effect of the operators in Eq.~(\ref{L_SILH}) the partial width of $h\to gg$ can be expressed as  \cite{Giudice:2007fh}
\be
\label{eq:onshellhgg}
\frac{\Gamma(h\to gg)}{\Gamma(h\to gg)_{SM}} \to  \left[ 1-\frac{v^2}{f^2} {\rm Re}\left( c_H + 2c_y - {6y_t^2 c_g}\, \frac{f^2}{m_\rho^2} \right) \right] \ .
\ee
In fact, Eq.~(\ref{eq:onshellhgg}) does not depend on the PNGB nature of the Higgs and can be applied to any models once $f$ and $m_\rho$ are properly identified. Throughout this work we define
\be
\label{eq:vdef}
v = \left(\sqrt{2}G_F\right)^{-\frac12} = 246 \quad {\rm GeV} \ .
\ee
As emphasized, the combination $c_H+2c_y$ appearing in Eq.~(\ref{eq:onshellhgg}) is independent of the particular operator basis one chooses. Therefore one could compute $c_H+2c_y$ in a general basis where $c_r\neq 0$ without explicitly going back to the SILH basis, which simplifies the computation, and then calculate the production rate using Eq.~(\ref{eq:onshellhgg}).

In the SILH lagrangian \cite{Giudice:2007fh}, the scale $f$ is identified with the ``pion decay constant'' in a non-linear sigma model (nl$\sigma$m) and $m_\rho$ with the mass scale of the composite resonances contributing to the loop-induced gluonic operator. Then the natural size of $c_H$, $c_y$, and $c_g$ is order unity, and all three operators must be included when computing the cross-section. To the contrary, in weakly-coupled theories where the Higgs appears as a fundamental particle at the TeV scale,  $f\to \infty$ and only the gluonic operator is important. Two such examples are  the weak-scale supersymmetry \cite{Dermisek:2007fi} and universal extra dimensions \cite{Petriello:2002uu}.

It may seem that, since  the effect of the compositeness is ${\cal O}(v^2/f^2)$, the deviation in the production rate would be in the order of a few percents for $f\agt 1$ TeV. However, it was shown in Ref.~\cite{Low:2009di} that all three coefficients in Eq.~(\ref{eq:onshellhgg}) tend to have the same sign that goes in the direction of reducing the overall production rate. Therefore, there could be ``pile-up'' effect so that the resulting change could be substantial. Indeed, we confirm this observation when computing the rate in explicit models in Section \ref{sect:models}.


\section{Production of a composite Higgs}
\label{sect:prod}

In this section we show  that the bulk of the calculation for $c_H$, $c_y$ and $c_g$ in a composite Higgs model can be done in full generality without referring to any specific models. We mostly follow the notations in and summarize the results of Ref.~\cite{Low:2009di}.

\subsection{$c_H$}

There are three possible sources for $c_H$:
\begin{itemize}

\item The nl$\sigma$m: 

Within the nl$\sigma$m the contribution to $c_H$ is most easily computed by observing that, turning on a background Higgs field $h^3 \to h^3+ h$, Eq.~(\ref{L_SILH}) gives rise to
 \be
 \label{eq:crch}
 {\cal L}_H \supset \frac14 \left[1+ \left(c_H+\frac{c_r}4\right)\frac{h^2}{f^2}\right]\partial_\mu h^3 \partial^\mu h^3 +\frac14 g^2 h^2 \left(1+\frac{c_r}4 \frac{h^2}{f^2}\right) W_\mu^+ W^{-\,\mu} \ ,
 \ee
 where we have included the effect of a non-vanishing ${\cal O}_r$ that is generally present in the nl$\sigma$m. However, it is easy to convince oneself that, using the CCWZ formulation of the nl$\sigma$m \cite{Coleman:sm,Callan:sn},  the kinetic term of the neutral Higgs boson is not corrected at any order in $h/f$ once the Goldstone kinetic term is canonically normalized at the leading order. We then conclude\footnote{This can also be derived using the fourth-rank tensor, ${\cal T}^{abcd}$, defined in Ref.~\cite{Low:2009di}.}
 \be
 \label{eq:nlsmbasis}
 c_H=-c_r/4 \ ,
 \ee
which is the ``natural'' basis for the nl$\sigma$m. On the other hand, we see that  $c_r$ can be computed from the SM $W$ boson mass in the presence of a background Higgs field:\footnote{At the level of dimension-six, $c_r$ is the only term contributing to the $W^\pm$ mass. See also footnote \ref{cusfoot}.}
\be
\label{eq:nlsmch}
m_W^2(h) = \frac14 g^2 h^2 \left(1-c_H^{(\sigma)} \frac{h^2}{f^2}\right) \ .
\ee
Although not obvious in this formulation, it was proven in Ref.~\cite{Low:2009di} that $c_H^{(\sigma)}> 0$.

\begin{figure}[t]
\includegraphics[scale=0.65]{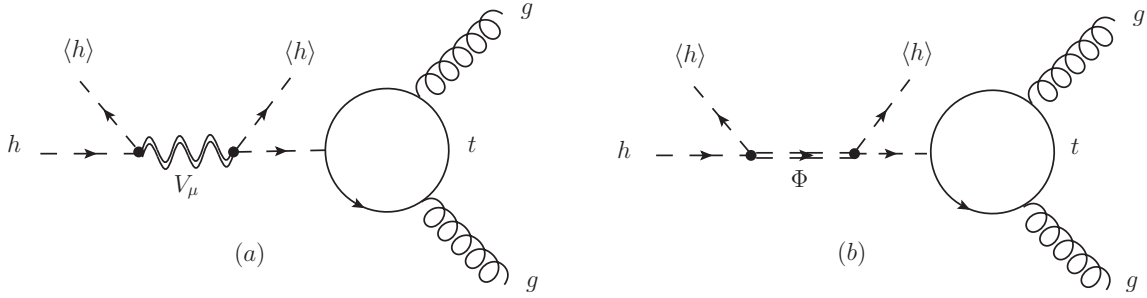} 
\caption{\em \label{fig:fig2} Feynman diagrams for contributions to the Higgs production from integrating out a heavy vector $V_\mu$ and a heavy scalar $\Phi$ are shown in $(a)$ and $(b)$, respectively. These contributions do not depend on the PNGB nature of the Higgs. }
\end{figure} 

\item Integrating out heavy vectors: 

A contribution to $c_H$ could follow from integrating out a heavy vector coupling to  the Higgs current. There are two types of currents which contribute:
\be
J_{HL\,\mu}^a =    i H^\dagger \, \frac{\sigma}{2}^a \tensor{D}_\mu \, H \ , \qquad
  J_{HR\,\mu}^-= i \, H^T\,\epsilon  \tensor{D}_\mu \,H\ ,
\ee
where $\epsilon=i\sigma^2$. Denoting the vector fields coupling to $J_{HL\,\mu}^a $ and $J_{HR\,\mu}^-$ by $V_\mu^a$ and $V_\mu^+$, respectively, we parametrize the relevant interactions as
\be
\frac{m_L^2}{2}V^{a\mu}V^a_\mu+ g_{\rho L} \gamma_H  V^{a\mu} J_{HL\mu}^a  +{m_+^2}V^{+\mu}V^-_\mu+\frac{g_{\rho R}\zeta_H}{\sqrt 2}\left ( V^{+\mu} J_{HR\mu}^-+{\rm h.c.}\right ) \ ,
\ee
from which we arrive at:
\be
\label{eq:vecchcy}
c_H^{(v)} = \frac{ g_{\rho L}^2 \gamma_H^2 f^2}{4 m_L^2}+ \frac{ g_{\rho R}^2 \zeta_H^2 f^2}{ m_+^2} >0\ .
\ee
See Fig.~2(a) for the diagrammatic contribution. It turns out that we also have $c_H^{(v)}=-c_r^{(v)}/4$ in this case. Notice that in deriving Eq.~(\ref{eq:vecchcy}) it is important to use a ``universal basis'' \cite{Barbieri:2004qk} where all new physics effects are described by oblique operators involving only vector bosons and Higgses, so that there are no corrections to vertices between SM vectors and fermions. This is equivalent to using the SM equations of motion to express the SM fermionic current in terms of operators involving only the Higgs and vector fields. Notice there is no contribution to $c_y$ here and Eq.~(\ref{eq:vecchcy}) is consistent with $(c_H+2c_y)|_{SILH}$ obtained from integrating out the vectors in Ref.~\cite{Low:2009di}.

\item Integrating out heavy scalars: 

The only possible quantum numbers for a heavy scalar coupling to two Higgses at the tree-level are a real electroweak singlet $\Phi_s$, a real electroweak triplet $\Phi_r^a$, and a complex electroweak triplet $\Phi_c^a$. However, a real triplet scalar does not contribute to $c_H$ \cite{Low:2009di}. If we parametrize the relevant interactions in the effective lagrangian as
\bea
&&-\frac12\Phi_s\Box\Phi_s-\frac12 m_s^2 \Phi_s^2 +
\beta_s f \, \Phi_s H^\dagger H \\
&& \qquad -\Phi_c^*\Box\Phi_c - m_c^2 |\Phi_c^a|^2 +\beta_c f \, 
\left( \Phi_c^a H^T\epsilon \frac{\sigma^a}2 H + {\rm h. c.}\right) \ ,
\eea
then we obtain
\be
\label{eq:scalarchcy}
c_H^{(s)} = + \frac{\beta_c^2 f^4}{2m_c^4}  + \frac{\beta_s^2 f^4}{m_s^4}>0 \ .
\ee
The diagrammatic contribution is shown in Fig. 2(b). Again Eq.~(\ref{eq:scalarchcy}) is consistent with $(c_H+2c_y)|_{SILH}$ obtained from integrating out the scalars in Ref.~\cite{Low:2009di}.

\end{itemize}

Note that all three contributions to $c_H$ are positive and go in the direction of reducing the production rate \cite{Low:2009di}.  Moreover, effects of integrating out heavy scalars and vectors are independent of the PNGB nature of the Higgs and would be there whenever there are heavy particles coupling to the Higgs at the tree-level, even if the Higgs is a fundamental scalar.

\subsection{$c_g$ and $c_y$}
\label{sect:Og}

The computation of $c_g$ from integrating out heavy colored particles is performed via the Higgs low-energy theorem \cite{Ellis:1975ap,Shifman:1979eb}, which instructs us to compute the contribution to the renormalization group running of the low-energy effective coupling constant due to the heavy colored particle in the presence of a non-zero background Higgs field $h$. The leading interaction between $h$ and the gluons is then obtained by expanding the $h$-dependent mass of the heavy particle in the one loop effective lagrangian:
\bea
\label{generalcoupling}
{\cal L}_{hgg}&=&\frac{g_s^2}{48\pi^2}\frac{h}{v}\left [ \sum_{r_F} t_{r_F} \frac {\partial}{\partial\log v}  \log \det \left({m}_{r_F}^\dagger(v) {m}_{r_F}(v)\right) \right.\nonumber \\
&&\left. \qquad\quad +\frac{1}{4}\sum_{r_S} t_{r_S}\frac {\partial}{\partial\log v} \log \det \left({m}_{r_S}^\dagger(v) {m}_{r_S}(v)\right)\right ]G_{\mu\nu}^a G^{a\, \mu\nu} \ ,
\eea
where the two sums are over Dirac fermions and complex scalars, respectively, while $t_r$ is the Dynkin index of the multiplet, which is $1/2$ for the fundamental representation, and ${m}(h)$ is the mass of heavy particle in the presence of the background Higgs field.  Two assumptions are made in Eq.~(\ref{generalcoupling}): i) the scalar Higgs $h$ has a VEV, $\langle h\rangle =v$, where $v$ is defined in Eq.~(\ref{eq:vdef}), and ii) $h$ has a canonical kinetic term. In the SILH basis condition i) is satisfied, while a rescaling $h\to h/\sqrt{1+c_{H} v^2/f^{2}}$ is required to bring the Higgs kinetic term back to canonical normalization.\footnote{\label{foot3} In a general basis where $c_r \neq 0$, the relation between $\langle h\rangle$ and  $v$ is derived from Eq.~(\ref{eq:crch}):
 \be
 v^2 = \langle h \rangle ^2 \left( 1+ \frac{c_r}4 \frac{\langle h \rangle ^2}{f^2} \right)\ . \nonumber
 \ee
 In the SILH basis $c_r=0$ and we have $\langle h\rangle = v$, while in a general basis the Higgs kinetic term needs to be re-scaled by $h\to h/\sqrt{1+(c_H+c_r/4) \langle h \rangle^2/f^2}$ and  $v = \langle h \rangle ( 1+ \frac{c_r}8 {\langle h \rangle ^2}/{f^2})$. So in the end we obtain the same result as in the SILH basis.}

For composite Higgs models we focus on the case of Dirac fermions in the fundamental representation of $SU(3)_c$. In a given model the fermion mass matrix in the top sector ${\cal M}(h)$ is often not diagonal, and to compute $c_g$ using Eq.~(\ref{generalcoupling}) one needs to solve for the mass eigenvalue of the heavy top partner. On the other hand,
the light mass eigenstate is taken to be the SM top quark.  Expanding with respect to the background field $h$, the top quark mass can be written as
\be
m_t(h) = \frac{\lambda_t}{\sqrt{2}}h\left(1-c_y^{(t)} \frac{h^2}{2f^2} \right) +{\cal O}(h^5) \ ,
\ee
which contains the correction to the top Yukawa coupling $c_y^{(t)}$ and must be included in the computation. However, the effects of $c_g$ and $c_y^{(t)}$ can be included at once if we use the full fermion mass matrix ${\cal M}(h)$ in Eq.~(\ref{generalcoupling}), without having to solve for the mass eigenvalues explicitly.  In other words,
\be
\label{eq:vhdiff}
\left. \frac{1}{2} \frac{\partial}{\partial \log  h}  \log \frac{{\cal M}(h)^2}{\mu^2}\right|_{h=\langle h \rangle} = 1-  c_y^{(t)} \frac{v^2}{f^2} + 3 y_t^2 c_g \frac{v^2}{m_\rho^2} \ ,
 \ee
 where we have neglected higher order terms in $v/f$.\footnote{As explained in footnote \ref{foot3}, there is a subtlety arising from the fact that, in a general basis where $c_r \neq 0$, the relation between $\langle h\rangle$ and  $v$  is non-trivial. However, this distinction in Eq.~(\ref{eq:vhdiff}) is only higher order in $v/f$ and we could use $\langle h \rangle$ and $v$ interchangeably.} This observation is particularly useful when there are multiple top partners and the mass matrix is difficult to diagonalize.

\vspace{0.5cm}

Summarizing the discussion so far, the effective on-shell coupling of the Higgs with two gluons is 
\be
\label{eq:hggexp}
 {\cal L}_{hgg} =  \frac{g_s^2}{48\pi^2}\frac{h}{v}\left[ \left. \frac{1}{2} \frac{\partial}{\partial \log  h}  \log \frac{{\cal M}(h)^2}{\mu^2}\right|_{h=v} -  \frac{c_H}2 \frac{v^2}{f^2}  \right]G_{\mu\nu}^a G^{a\, \mu\nu}  \ ,
\ee
where $c_H =c_H^{(\sigma)}+c_H^{(s)}+c_H^{(v)}$. On the other hand, the SM contribution is 
\be
 {\cal L}_{hgg}^{(SM)} =  \frac{g_s^2}{48\pi^2}\frac{h}{v}\, G_{\mu\nu}^a G^{a\, \mu\nu} \ .
 \ee
A comparison of the production rates could be obtained by squaring the coefficients of the $h G_{\mu\nu}^a G^{a\, \mu\nu} $ operator.


\section{Explicit Models}
\label{sect:models}

In this section we consider two classes of explicit examples: the little Higgs theories \cite{Arkani-Hamed:2001nc, Arkani-Hamed:2002qx, Arkani-Hamed:2002qy}, which employs the collective breaking mechanism, and the holographic Higgs theories \cite{Contino:2003ve, Agashe:2004rs}. In the first class we compute the production rate in the $SU(5)/SO(5)$ littlest Higgs  \cite{Arkani-Hamed:2002qy} and its T-parity version \cite{Cheng:2003ju, Pappadopulo:2010jx}, as well as the $SO(9)/SO(5)\times SO(4)$ littlest Higgs with a custodial symmetry \cite{Chang:2003zn}, while in the second class we discuss  two different implementations of the top sector \cite{Contino:2006qr}, which protect the $Zbb$ vertex with a custodial symmetry \cite{Agashe:2006at}, in the $SO(5)/SO(4)$ minimal composite Higgs model. We will refer the detail of each model to the original construction.

\subsection{The $SU(5)/SO(5)$ littlest Higgs}

The littlest Higgs \cite{Arkani-Hamed:2002qy}, based on the $SU(5)/SO(5)$ coset, is one of the most popular little Higgs models. Because of the lack of the custodial symmetry, the original construction is severely constrained by the precision electroweak measurements \cite{Csaki:2002qg}. However, the constraint could be relaxed significantly if one only gauges one $U(1)$ group \cite{Csaki:2003si}, which leaves an extra singlet scalar in the PNGB matrix. This is the version we consider below.
The PNGB matrix has the form 
\beq
\label{eq:LHpi}
\Pi=\begin{pmatrix}
 \frac{1}{2\sqrt 5} \eta_s &   H & \Phi \\ H^\dagger & -\frac{2}{\sqrt 5}\eta_s&  H^T\\
\Phi^\dagger &  H^* &  \frac{1} {2\sqrt 5} \eta_s 
\end{pmatrix},
\eeq
where $\Phi$ is a complex triplet with hypercharge $Y=+1$,  $H$ is the Higgs doublet with $Y=+1/2$, and $\eta_s$ is a singlet scalar without hypercharge\footnote{It has been pointed out in Ref.~\cite{Schmaltz:2008vd} that a real singlet induces a quadratically divergent Higgs mass at one-loop level. Thus the Higgs mass in this model is really fine-tuned.}:
\beq
\Phi=\begin{pmatrix} \Phi^{++} & \frac{\Phi^+}{\sqrt2}\\ \frac{\Phi^+}{\sqrt2} &  \Phi^0 \end{pmatrix},\qquad H=
\begin{pmatrix} H_1 \\ H_2 \end{pmatrix}= \frac1{\sqrt2}\begin{pmatrix}h_1+i h_2\\ h_3+ih_4 \end{pmatrix} \ .
\eeq
The Higgs VEV is $\left<H\right>=(0, v)^T/\sqrt{2}$. The nl$\sigma$m field $\Sigma(x)$ has the form
\be
\label{eq:sigmalh}
\Sigma(x) = e^{i\Pi/f}\Sigma_0 e^{i\Pi/f} = e^{2i\Pi/f}\Sigma_0 \ , \qquad 
\Sigma_0 = \begin{pmatrix}
              \phantom{xxx}  & & \openone_2 \\
                & 1 & \phantom{xxx} \\
              \openone_2 & & 
            \end{pmatrix} \ ,
\ee
where $\openone_2$ is a $2\times 2$ unit matrix. The gauge generators are defined as follows:
\bea
Q_1^a&=&\begin{pmatrix}
\sigma^a/2 &  \phantom{XX} \\
    &   
\end{pmatrix}\ ,\qquad 
Q_2^a=\begin{pmatrix}
  & \\
\phantom{xx} & -\sigma^{a*}/2 
\end{pmatrix}\ , \\
 Y&=&\textrm{diag}\left(\frac{1}{2},\frac{1}{2},0,-\frac{1}{2},-\frac{1}{2}\right)\ .
\eea
The kinetic term can be written as
\bea
 \mathcal L&=&\frac{f^2}{8}\text{Tr}[ D_\mu\Sigma^\dagger  D^\mu\Sigma] \ , \\
  D_\mu \Sigma &=& \partial_\mu \Sigma -i \sum_j\left[ g_j W_j^a \left(Q_j^a \Sigma+\Sigma Q_j^{a\, T}\right)\right] 
   -ig^\prime B_\mu (Y\Sigma+\Sigma Y) \ ,
\eea
where $g_j$ is the coupling of the $[SU(2)]_j$ group and $g^\prime$ that of the $U(1)_Y$ group.

\subsubsection{$c_H$}

The SM $W$ boson mass $m_W(h)$ can be computed using the kinetic term, 
\be
m_W^2(h) = \frac12 g^2 f^2 \sin^2 \frac{h}{\sqrt{2}f} \ , \qquad g^2 = \frac{g_1^2 g_2^2}{g_1^2+g_2^2} \ ,\\
\ee
from which $c_H$ follows by Eq.~(\ref{eq:nlsmch}):
\be
\label{eq:LLHchsi}
c_H^{(\sigma)}=   \frac16 \  .
\ee
Since the SM fermions are taken to transform under only one of the $SU(2)$ gauge group, say the upper $[SU(2)]_1$, we identify the low-energy interpolating field for the SM gauge fields as $W^a_\mu\equiv A_{1\mu}^a$, which will be kept fixed when integrating out the heavy gauge fields. Notice that $W^a_\mu$ is not a mass eigenstate. The interpolating field for the heavy combination is taken to be the mass eigenstate: $W^{\prime\,a}\equiv (g_1 A^a_1-g_2 A^a_2)/{\sqrt{g_1^2+g_2^2}}$. Then it is straightforward to compute
\be
m_{W^\prime} = g_{\rho L} f \ , \qquad g_{\rho L} \equiv \frac12 \sqrt{g_1^2+g_2^2}\ , \qquad {\rm and} \qquad 
\gamma_H  = 1\ ,
\ee
which leads to 
\be
\label{eq:LLHchv}
c_H^{(v)} = \frac14 \ .
\ee
A general and detailed implementation of this procedure can be found in the appendix B of Ref.~\cite{Low:2009di}. On the other hand, the triplet scalar potential is given by 
\beq
\label{eq:scalarp}
 V(\Phi,\Phi^*) = c_+ f^2 \left| \Phi_{ij} + \frac{i}{2f}(H_i H_j+H_j H_i)\right|^2 
   + c_- f^2 \left| \Phi_{ij} - \frac{i}{2f}(H_i H_j+H_j H_i)\right|^2 ,
   \eeq
which gives rise to
\be
\label{eq:LLHchs}
c_H^{(s)}=\frac{(c_+ -c_-)^2}{(c_+ +c_-)^2} \ . 
\ee
Notice that the singlet scalar $\eta_s$, as it stands now, is an exact Goldstone boson. Additional source of symmetry breaking must be present in order to give $\eta_s$ a potential, which at the same time would also contribute a mass to the Higgs boson. Then stability of the Higgs mass would require $\eta_s$ to be lighter than the Higgs \cite{Surujon:2010ed}. Thus we do not consider the effect of the singlet scalar.

We should also comment that in general the triplet receives a non-vanishing VEV, $v'$,  which is severely constrained by the precision electroweak data. The effect of $v'$ can be reproduced within the effective lagrangian by adding to Eq.~(\ref{L_SILH})
\be
\frac{4v^{\prime\,2}}{v^4} H^\dagger H(D_\mu H^\dagger D^\mu H)-\frac{v^{\prime\,2}}{v^4} \left|H^\dagger\tensor{D}_\mu H\right|^2 \ .
\ee
However, there is no contribution to $c_H+2c_y$ from the triplet VEV.

\subsubsection{$c_g$ and $c_y$}

The top Yukawa sector of the littlest Higgs has the form
\be
\mathcal L_{t}= {\lambda_1f} \epsilon_{ijk} \epsilon_{ab} \chi_i \Sigma_{ja}\Sigma_{kb} u^c + \lambda_2 f U U^c \,
\ee
where $\chi=(b\  t\  U)$, $i,j, k=1,2,3$, and $x,y=4,5$. From the Yukawa interactions we can work out the fermion mass matrix in the basis
\be
(u^c \ U^c) {\cal M} \begin{pmatrix}  t \\
                                                               U \end{pmatrix} ,
\ee                                                               
from which we compute the determinant of the mass matrix squared:
\bea
{\rm Det}\left({\cal M}^\dagger{\cal M}\right) &=& 2\lambda_1^2 \lambda_2^2\, f^4 \sin^2\frac{\sqrt{2}v}{f} \ , \\
\label{eq:LLHfm}
\frac12 v \frac{\partial}{\partial v}\log {\rm Det}\left({\cal M}^\dagger{\cal M}\right)  &=& 1-\frac23 \frac{v^2}{f^2} \ .
\eea
In the above we have set the triplet VEV $v'=0$, whose effect is higher order in $(v/f)^2$, since the positivity of the triplet mass requires $v'/v < v/(4f)$ \cite{Csaki:2002qg}.

\begin{figure}[t]
\includegraphics[scale=0.85]{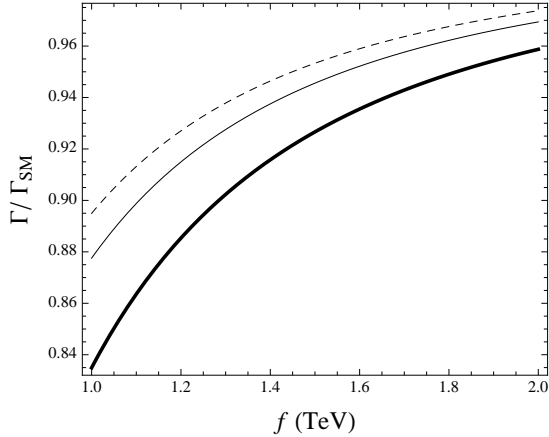} 
\caption{\em The ratio of the Higgs production rates in the gluon fusion channel in the littlest Higgs model over the SM expectation. The three curves, from top to bottom, are for $c_-/c_+= 1$, $0.3$,  and $0$, respectively.  Precision electroweak constraints require $f \agt$ $1.2$ TeV. \label{fig:fig3} }
\end{figure} 

Putting Eqs.~(\ref{eq:LLHchsi}), (\ref{eq:LLHchv}), (\ref{eq:LLHchs}), and  (\ref{eq:LLHfm}) in Eq.~(\ref{eq:hggexp}) we obtain
\be
\label{eq:lhggh}
 {\cal L}_{hgg} =  \frac{g_s^2}{48\pi^2}\frac{h}{v}\left[1-\left(\frac{7}{8}+\frac{(c_+-c_-)^2}{2(c_++c_-)^2}\right)\frac{v^2}{f^2}  \right]G_{\mu\nu}^a G^{a\, \mu\nu} \ .
 \ee
Therefore for $f\sim 1.2$ TeV, which is allowed by the precision electroweak constraint and corresponds to the limit  $c_+\gg c_-$ \cite{Csaki:2003si}, the reduction from the SM expectation is at ${\cal O}(12\%)$. This number is slightly larger than the previous number obtained in Ref.~\cite{Han:2003gf}, where some of the effects were neglected.  In Fig.~\ref{fig:fig3} we show the deviation from the standard model expectation for three different choices of $c_-/c_+$. One surprising feature of Eq.~(\ref{eq:lhggh}) is that the mass of the heavy top partner does not enter into the production rate at all, in sharp contrast to the MSSM and the UED models, where the rate is very sensitive to the stop mass  \cite{Dermisek:2007fi} as well as the 1st KK top mass \cite{Petriello:2002uu}. We will see that this feature persists in other types of composite Higgs models.

\subsection{The (new) littlest Higgs with T-parity}

Implementation of T-parity in the littlest Higgs model was initiated in Ref.~\cite{Cheng:2003ju} and completed recently in Ref.~\cite{Pappadopulo:2010jx}, which model we focus here. The model is based on the coset
\be
\frac{SU(5)}{SO(5)} \times \frac{[SU(2)\times U(1)]_L\times [SU(2)\times U(1)]_R}{[SU(2)\times U(1)]_V} \ ,
\ee
which contains two link fields $\Sigma$ and $X$. The Higgs doublet is contained in $\Sigma$, which is identical to that defined in Eq.~(\ref{eq:sigmalh}) in the original littlest Higgs model, while $X$ does not enter into our calculation here.  Results in this subsection are new.

\subsubsection{$c_H$}

Only the Higgs doublet is even under T-parity while all other scalars, as well as the heavy vector bosons, are T-odd. Thus there is no tree-level interaction between two Higgs scalars and one single heavy scalars/vectors, which implies the only contribution to $c_H$ comes from that of the nl$\sigma$m. Since the part of the coset involving the Higgs is identical to the original littlest Higgs, we only need to quote from Eq.~(\ref{eq:LLHchsi}):
\be
c_H=   \frac16 \  .
\ee
 
 \subsubsection{$c_g$ and $c_y$}
 
 The top sector of the model embeds two doublets $\psi_{1,2}$ and two singlets $\chi_{1,2}$ into incomplete representations of $SU(5)$, a $\mathbf{5}$ and a $\mathbf{5}^*$:
\be
{\cal T}_1 =\begin{pmatrix} 
                                  \psi_1 \\
                                  \chi_1 \\
                                  0
                                  \end{pmatrix} \ , \qquad
 {\cal T}_2 =\begin{pmatrix} 
                                 0 \\
                                  \chi_2 \\
                                   \psi_2
                                  \end{pmatrix} \ .
\ee                                  
Under T-parity we have $\psi_1 \leftrightarrow -\psi_2$ and $\chi_1 \leftrightarrow \chi_2$.  The Yukawa coupling arises from the following lagrangian:
\be
\frac{\lambda_1 f}4 \epsilon_{ij}\epsilon_{abc} \left(\Sigma_{ai}\Sigma_{bj} \overline{\cal T}_{1\, c}  + \mathbf{T}[ \Sigma_{ai}\Sigma_{bj} \overline{\cal T}_{1\, c} ] \right) t_R + \frac{\lambda_2 f}{\sqrt{2}}(\overline{\chi}_1 \tau_1+\overline{\chi}_2 \tau_2) \ ,
\ee
where $\mathbf{T}[{\cal O}]$ is the image of ${\cal O}$ under T-parity, and $\tau_{1,2}$ are two additional singlets transforming under T-parity as $\tau_1 \leftrightarrow \tau_2$. The T-parity eigenstates are
\be
\frac{\psi_1-\psi_2}{\sqrt{2}} = \sigma^2 \begin{pmatrix} 
                                                            t_L \\
                                                            b_L 
                                                            \end{pmatrix} \ , \quad
\chi^{\pm} =    \frac{\chi_1\pm\chi_2}{\sqrt{2}}    \ , \quad
\tau^{\pm} =    \frac{\tau_1\pm\tau_2}{\sqrt{2}}   \ .
\ee                                                    
In the end only the T-even partners of the SM top participate in the cancellation of quadratic divergences, and the mass matrix is defined in the basis
\be
 ( \bar{t}_L , \bar{\chi}^+) {\cal M} 
                                                           \begin{pmatrix} 
                                                            t_R \\
                                                            \tau^+ 
                                                            \end{pmatrix}   \ ,
\ee                                                                                                                 
from which we compute
\bea
{\rm Det}\left({\cal M}^\dagger{\cal M}\right) &=& \frac18 \lambda_1^2 \lambda_2^2\, f^4 \sin^2\frac{\sqrt{2}v}{f} \ , \\
\label{eq:LLHfm}
\frac12 v \frac{\partial}{\partial v}\log {\rm Det}\left({\cal M}^\dagger{\cal M}\right)  &=& 1-\frac23 \frac{v^2}{f^2} \ .
\eea

\begin{figure}[t]
\includegraphics[scale=0.85]{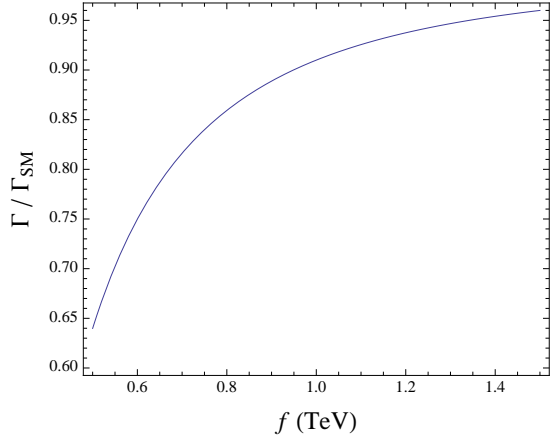} 
\caption{\em The ratio of the Higgs production rates in the gluon fusion channel in the (new) littlest Higgs model with T-parity \cite{Pappadopulo:2010jx} over the SM expectation. Precision electroweak constraints is quite weak and allow  $f \agt 500$ GeV. \label{fig:figtp} }
\end{figure}

Combining the results we obtain
\be
\label{eq:lhgghtp}
 {\cal L}_{hgg} =  \frac{g_s^2}{48\pi^2}\frac{h}{v}\left[1-\frac{3}{4}\frac{v^2}{f^2}  \right]G_{\mu\nu}^a G^{a\, \mu\nu} \ .
 \ee
 The constraint on $f$ is very weak in models with T-parity. In this model it was found that $f\agt 500$ GeV is still allowed \cite{Pappadopulo:2010jx}, at which value the reduction could approach 35\%.  In Fig.~\ref{fig:figtp} we plot the ratio of the predicted production rate in the T-parity scenario over the standard model versus $f$.

\subsection{The $SO(9)/SO(5)\times SO(4)$ littlest Higgs with custodial symmetry}

This model alleviates the constraint from the electroweak $\rho$ parameter with the help of the custodial symmetry. We use the same basis for the $SO(9)$ generators, as well as the $SU(2)_L\times SU(2)_R \cong SO(4)$ generators, as in Ref.~\cite{Chang:2003zn}. The structure of the model is similar to that of the $SU(5)/SO(5)$ model, with the nl$\sigma$ field $\Sigma(x)$:
\be
\Sigma(x) = e^{i\Pi/f}\Sigma_0 e^{i\Pi/f} = e^{2i\Pi/f}\Sigma_0 \ , \qquad 
\Sigma_0 = \begin{pmatrix}
              \phantom{xxx}  & & \openone_4 \\
                & 1 & \phantom{xxx} \\
              \openone_4 & & 
            \end{pmatrix} \ ,
\ee
where $\openone_4$ is a $4\times 4$ unit matrix. The PNGB matrix contains the following scalars transforming under $SU(2)_L\times SU(2)_R$,
\be
h:(\mathbf{2}_L,\mathbf{2}_R) \qquad \phi^0 :(\mathbf{1}_L,\mathbf{1}_R) \qquad \phi^{ab}:(\mathbf{3}_L,\mathbf{3}_R) 
\ee
and is given by
\beq
\label{eq:LHcuspi}
\Pi=\frac{-i}{4} \begin{pmatrix}
 0 &   \sqrt{2} \vec{h} & -\Phi \\ -\sqrt{2}\vec{h}^T  &0 &  \sqrt{2}\vec{h}^T\\
\Phi&  -\sqrt{2} \vec{h} & 0
\end{pmatrix}\ ,
\eeq
where $\vec{h}=( -h^2,   -h^1 , -h^4,  h^3)$ and $ \Phi= \phi^0 +4\phi^{ab} T^{la} T^{rb}$. Here $T^{la}$ and $T^{rb}$ are the generators for the $SU(2)_L$ and $SU(2)_R$, respectively.     

Inside the $SO(9)$ the following subgroups are gauged:
\bea
SO(4)\cong SU(2)_L\times SU(2)_R &:& \tau^{la} = 
   \begin{pmatrix} T^{la} &  \\
                                           & 0_5 \end{pmatrix} \qquad 
                                            \tau^{ra} = 
   \begin{pmatrix} T^{ra} &  \\
                                           & 0_5 \end{pmatrix} \ , \nonumber \\
 SU(2)\times U(1) &:&  \eta^{la} = 
   \begin{pmatrix} 
      0_5 & \\
         & T^{la} \end{pmatrix} \qquad 
          \eta^{r3} = 
   \begin{pmatrix} 
      0_5 & \\
         & T^{r3} \end{pmatrix} \ . \nonumber 
\eea         
 The kinetic energy of the Goldstone bosons is
 \bea
        {\cal L} &=& \frac{f^2}4 {\rm Tr}[D_\mu\Sigma D^\mu\Sigma] \ , \qquad D_\mu\Sigma = \partial_\mu\Sigma + i [ A_\mu, \Sigma] \ , \\
        A_\mu &=& g_L W^{la}_{SO(4)} \tau^{la} + g_R W^{ra}_{SO(4)} \tau^{ra} + g_2 W^{la}\eta^{la} + g_1 W^{r3}\eta^{r3} \ .
\eea

\subsubsection{$c_H$}
The mass of the SM $W$ boson in this model are given by
\be
m_W^2(h)= g^2 f^2 \sin^2 \frac{h}{2f} \ , \qquad g^2= \frac{g_2^2 g_L^2}{g_2^2+g_L^2} \ , 
\ee
leading to
\be
\label{eq:LLHcuschsi}
c_H^{(\sigma)} = \frac1{12} \ .
\ee
Again the SM fermions are assumed to transform only under $SU(2)\times U(1)$, implying we need to hold fixed $W^{la}$ and $W^{r3}$ when integrating out the heavy mass eigenstates, which are taken to be
\be
B'=\frac{g_1 W^{r3} -g_R W^{r3}_{SO(4)}}{\sqrt{g_1^2+g_R^2}} \ , \quad W^{\prime\, a} = \frac{g_2 W^{la} -g_L W^{la}_{SO(4)}}{\sqrt{g_2^2+g_L^2}} \ ,\quad
W^{r\pm} = \frac{W^{r1}_{SO(4)} \mp i W^{r2}_{SO(4)}}{\sqrt{2}}\ .
\ee
The masses of the heavy gauge bosons are 
\be
m_{B'} = \sqrt{g_1^2+g_R^2}\, f\equiv g_{\rho B} f \ , \quad m_{W'} = \sqrt{g_2^2+g_L^2}\, f\equiv g_{\rho L} f \ , \quad m_{W^{r\pm}} = g_R\, f \equiv g_{\rho R} f\ ,
\ee
from which we compute $\gamma_H = 1/2$, $\zeta_H = 1/4$, and
\be
\label{eq:LLHcuschv}
c_H^{(v)}= \frac18 \ .
\ee
The scalar potential in this model is given by 
\bea
&& \lambda_{\mathbf{1}}^- f^2 (\phi^0 - {\cal H}^0)^2+\lambda_{\mathbf{1}}^+ f^2 (\phi^0 + {\cal H}^0)^2  \nonumber \\
&&\qquad+  \lambda_{\mathbf{3}}^- f^2 (\phi^{ab} - {\cal H}^{ab})^2+\lambda_{\mathbf{3}}^+ f^2 (\phi^{ab} + {\cal H}^{ab})^2 + \Delta\lambda_{\mathbf{3}} \left(\phi^{a3}+{\cal H}^{a3}\right)^2 \ ,
\eea
where ${\cal H}^0=|\vec{h}|^2/(2f)$ and ${\cal H}^{ab}=[(h^ch^c-h^4h^4)\delta^{ab}-2h^ah^b-2\epsilon^{abc}h^ch^4]/(8f)$. The contribution to $c_H$ is
\be
\label{eq:LLHcuschs}
c_H^{(s)}=\frac{(\lambda_{\mathbf{1}}^- -\lambda_{\mathbf{1}}^+)^2}{4(\lambda_{\mathbf{1}}^- +\lambda_{\mathbf{1}}^+)^2}+ \frac{(\lambda_{\mathbf{3}}^- -\lambda_{\mathbf{3}}^+)^2}{16(\lambda_{\mathbf{3}}^- +\lambda_{\mathbf{3}}^+)^2} \ .
\ee

\subsubsection{$c_g$ and $c_y$}

\begin{figure}[t]
\includegraphics[scale=0.85]{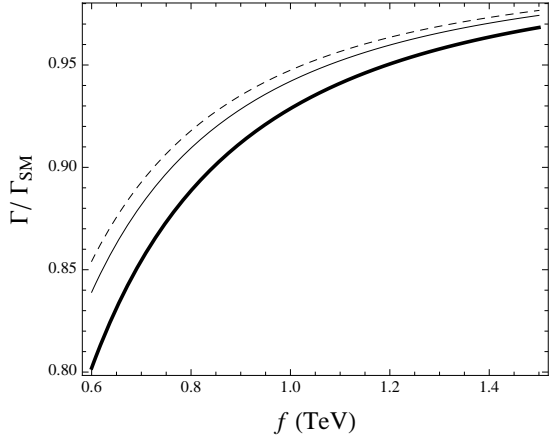} 
\caption{\em The ratio of the Higgs production rates in the gluon fusion channel in the littlest Higgs model with the custodial symmetry over the SM expectation. The three curves, from top to bottom, are for $\lambda_{\mathbf{1}}^- /\lambda_{\mathbf{1}}^+ =\lambda_{\mathbf{3}}^- /\lambda_{\mathbf{3}}^+= 1$, $0.3$,  and $0$, respectively.  Precision electroweak constraints require $f \agt$ $700$ GeV. \label{fig:fig4} }
\end{figure} 

In the top sector one introduces one $SO(4)$ vector $\vec{\cal X}$ which includes two $SU(2)_L$ doublets, ${\cal X}_1=(U_1, \psi_1)^T$ and ${\cal X}_2=(\psi_2, U_2)^T$, with hypercharges 1/6 and 7/6, respectively. An additional top-like singlet $t^c$ is also introduced to combine with $\vec{\cal X}^c$ to form a $SO(5)$ vector. The top Yukawa coupling is then
\be
{\cal L}_t = y_1 f (\vec{\cal X}^{c\, T}, t^c, 0_4) \Sigma 
     \begin{pmatrix} 0_5 \\
        \vec{q}_t  \end{pmatrix} + y_2 f \vec{\cal X}^T \vec{\cal X}^c + {\rm h. c. } \ ,
\ee
where $\vec{q}_t$ is the third generation quark doublet $Q_3=(t, b)^T$ written in the $SO(4)$ notation, as explained in the appendix of Ref.~\cite{Chang:2003zn}.  There are six top-like quarks in the construction, whose mass matrix is defined as
\be
(t, U_1, U_2) {\cal M}\begin{pmatrix}
              t^c \\
              U_1^c \\
              U_2^c 
            \end{pmatrix} \ .
\ee            
The determinant of the mass matrix-squared can be computed:
\bea
{\rm Det}\left({\cal M}^\dagger{\cal M}\right) &=& \frac14 y_1^2 y_2^4\, f^6 \sin^2\frac{v}{f} \ , \\
\label{eq:LLHcusfm}
\frac12 v \frac{\partial}{\partial v}\log {\rm Det}\left({\cal M}^\dagger{\cal M}\right)  &=& 1-\frac13 \frac{v^2}{f^2} \ .
\eea
         
Putting everything together we have
\be
 {\cal L}_{hgg} =  \frac{g_s^2}{48\pi^2}\frac{h}{v}\left[1-\left(\frac{7}{16}+\frac{(\lambda_{\mathbf{1}}^+ -\lambda_{\mathbf{1}}^-)^2}{8(\lambda_{\mathbf{1}}^+ +\lambda_{\mathbf{1}}^-)^2}+ \frac{(\lambda_{\mathbf{3}}^+ -\lambda_{\mathbf{3}}^-)^2}{32(\lambda_{\mathbf{3}}^+ +\lambda_{\mathbf{3}}^-)^2} \right)\frac{v^2}{f^2}  \right]G_{\mu\nu}^a G^{a\, \mu\nu} \ .
 \ee
 Because of the custodial symmetry, the electroweak constraint on this model is quite weak, with the $f$ being as low as 700 GeV \cite{Chang:2003zn}. In this case, the reduction from the SM expectation is 15\% in the limit $\lambda_{\mathbf{1}}^- \gg \lambda_{\mathbf{1}}^+$ and $\lambda_{\mathbf{3}}^- \gg \lambda_{\mathbf{3}}^+$. In Fig.~\ref{fig:fig4} we again show the prediction in the custodial littlest Higgs model over the standard model rate, by choosing $\lambda_{\mathbf{1}}^- /\lambda_{\mathbf{1}}^+ =\lambda_{\mathbf{3}}^- /\lambda_{\mathbf{3}}^+= 1$, $0.3$,  and $0$. The gluon fusion production of the Higgs in this model has not be computed previously.

\subsection{The $SO(5)/SO(4)$ minimal composite Higgs model}

The nl$\sigma$m field is parametrized by 
\be
\Sigma= \Sigma_0 e^{i\Pi/f} \ , \quad \Sigma_0 = (0,0,0,0,1) \ , \quad \Pi= i \begin{pmatrix}
   0_4 & -\vec{h}\\
   \vec{h}^T & 0 \end{pmatrix}\ .
    \ee
The kinetic term is simply
\be
\frac{f^2}{2} (D_\mu\Sigma) (D^\mu\Sigma) \ .
\ee    
Below the mass scale of the CFT resonances, $m_\rho = g_\rho f$, the spectrum consists of only the SM particles and the Higgs boson. However, in order to satisfy the $Zbb$ constraint, the top sector requires specific choices of $SO(5)$ representations. Following Ref.~\cite{Contino:2006qr}, we will consider two such choices, where the fermions in the top sector fill out a $\mathbf{5}$ (MCHM$_5$) and  a $\mathbf{10}$ (MCHM$_{10}$) of $SO(5)$, respectively. In these cases, it was found that there exist anomalously light fermionic states below $m_\rho$, whose effect should be considered within the effective lagrangian.

\subsubsection{$c_H$}

The $W$ mass is given by
\be
m_W^2(h) = \frac14 g^2 f^2 \sin^2\frac{h}{f} \ ,
\ee
from which we obtain
\be
c_H^{(\sigma)} = \frac13 \ .
\ee
Since the $SO(5)/SO(4)$ coset describes interactions below the scale $m_\rho$, which contains only the SM particles, the Higgs, and the anomalously light fermionic resonances, it is not necessary to include effects of integrating out spin-1 resonances in the CFT, whose masses are at or heavier than $m_\rho$.\footnote{We are grateful to R. Rattazzi for pointing this out to us.}

\subsubsection{$c_g$ and $c_y$}

The effective lagrangian in the top sector can be constructed following the philosophy of Ref.~\cite{Contino:2006nn}, by considering a weakly-coupled sector of elementary fermions $q_L=(t_L, b_L)$ and $t_R$, and a composite sector comprising the Higgs as well as fermions transforming under full representations of $SO(5)$. The composite fermions are $\Psi_L$ and $\Psi_R$ sitting in the $\mathbf{5}_{2/3}$ and $\mathbf{10}_{2/3}$ of $SO(5)\times U(1)_X$ in MCHM$_5$ and MCHM$_{10}$, respectively. Note the hypercharge $Y=T_R^3+X$. Furthermore, the Higgs only couples to the composite fermions in an $SO(5)$ invariant fashion. The composite and elementary sectors are linearly coupled only through the mass mixing term between the elementary and composite fermions, upon which the SM Yukawa interactions arise. In the end, the mass eigenstates, and hence the SM fermions, are linear combinations of elementary and composite states. The SM Yukawa coupling obviously is proportional to the ``partial compositeness'' of the SM field, resulting in the top quark being very composite, which is the reason for the lightness of its partners within the same $SO(5)$ multiplet \cite{Contino:2006qr}. 

Schematically, the effective lagrangian looks like
\be
{\cal L}_{\rm top} = {\cal L}_{\Psi} + {\cal L}_{\rm mix} \ ,
\ee
where ${\cal L}_{\Psi}$ is the $SO(5)$ invariant Yukawa coupling of $\Psi_L$ and $\Psi_R$ with the Higgs. The MCHM is equivalent to a 5D theory based on the AdS$_5$ space, according to the AdS/CFT correspondence \cite{Maldacena:1997re}. From the 5D perspective, $\Psi_L$ and $\Psi_R$ are the left-handed and right-handed chirality of one single 5D fermion and ${\cal L}_\Psi$ arises from the gauge-covariant derivative of the 5D fermion when the Higgs is identified with the $A_5$ component of the 5D gauge field \cite{Agashe:2004rs}. On the other hand, the mass mixing term is
\be
\label{eq:mix}
{\cal L}_{\rm mix}= \lambda_1 f \bar{q}_L Q_R +  \lambda_2 f \bar{T}_L t_R\ .
\ee
In the above $Q_R$ and $T_L$ are components of $\Psi_R$ and $\Psi_L$ with the appropriate quantum numbers. 

Now we are ready to write down the effective Yukawa couplings. 
\begin{itemize}

\item MCHM$_5$: $\Psi_L$ includes one $SU(2)_L$ doublets $q'_{L}=(t'_{L},b'_{L}$), one singlet $T_L$ and one exotic doublet $\tilde{Q}_L=(\chi_L, \tilde{T}_L)$ with hypercharge 7/6, while $\Psi_R$ contains one singlets $t'_{R}$, one doublet $Q_R=(T_R,B_R)$, and one exotic doublet $\tilde{Q}_R=(\chi_R, \tilde{T}_R)$. Notice that $\chi_R$ has the electric charge $Q=+5/3$. These composite fermions are embedded into $\mathbf{5}$'s, using the convention of $SO(5)$ matrices in \cite{Low:2009di},  as follows:
\be
\label{eq:psiL}
\Psi_L=\frac1{\sqrt{2}}\begin{pmatrix}
     b'_L +\chi_L \\
     i(b'_L-\chi_L) \\
     (t'_{L}+\tilde{T}_L) \\
     -i(t'_{L}-\tilde{T}_L) \\
     \sqrt{2} T_L \end{pmatrix} \ , \qquad
\Psi_R=\frac1{\sqrt{2}}\begin{pmatrix}
     B_R +\chi_R \\
     i(B_R-\chi_R) \\
     (T_R+\tilde{T}_R) \\
     -i(T_R-\tilde{T}_R) \\
     \sqrt{2} t'_{R} \end{pmatrix}     \ .     
\ee     
The top Yukawa coupling is written as\footnote{See alternative constructions of the effective top sector in Refs.~\cite{Falkowski:2007hz,Barbieri:2007bh}.}
\be
{\cal L}_{\mathbf{5}} =  \lambda_m \bar{\Psi}_L \Psi_R +\lambda_y (\bar{\Psi}_L \Sigma^T) (\Sigma \Psi_R) + {\cal L}_{\rm mix}+ {\rm h. c.}  \ ,
\ee
where ${\cal L}_{\rm mix}$ is the mixing between elementary fermions, $q_L=(t_L, b_L)$ and $t_R$, and the composite fermions, as defined in Eq.~(\ref{eq:mix}). Given the $U(1)_X$ charge assignment, $X=2/3$, there are four top-like Dirac fermions in MCHM$_5$: $({t}'_L, {T}_L, {t}_L, {\tilde{T}}_L)$ and $(T_R,\tilde{T}_R,  t'_R, t_R)$, leading to
\bea
{\rm Det}\left({\cal M}^\dagger{\cal M}\right) &=& \frac18 \lambda_1^2 \lambda_2^2 \lambda_y^2\lambda_m^2 \, f^8 \sin^2\frac{2v}{f} \ , \\
\label{eq:mchm5fm}
\frac12 v \frac{\partial}{\partial v}\log {\rm Det}\left({\cal M}^\dagger{\cal M}\right)  &=& 1-\frac43 \frac{v^2}{f^2} \ .
\eea
This result is the same as that coming from an alternative construction of the top Yukawa coupling in Ref.~\cite{Falkowski:2007hz}, which also takes into account the effect of $c_H$ correctly. The same rate has also been computed in Ref.~\cite{Espinosa:2010vn}, whose results we agree with.

It is perhaps worth commenting that the rate computed in  Ref.~\cite{Espinosa:2010vn} ignored the contribution of the fermionic top partners, which could have sizable couplings to the Higgs boson and thus modify the Higgs-gluon-gluon vertex significantly when the top partners are light. Our results here justify this approximation {\em a posteriori}.

\item MCHM$_{10}$: The $\mathbf{10}_{2/3}$ of $SO(5)\times U(1)_X$ can be decomposed into $\mathbf{4}\oplus\mathbf{6}$ under the $SO(4)$.  The $\mathbf{6}$ of $SO(4)$ further decomposes into $(\mathbf{3}_L, \mathbf{1}_R)$ and $(\mathbf{1}_L, \mathbf{3}_R)$ under $SU(2)_L\times SU(2)_R$. Since the electric charge $Q=T_L^3+Y=T_L^3+T_R^3+X$, it is simple to see that there is one top-like fermion in each of $(\mathbf{3}_L, \mathbf{1}_R)$ and $(\mathbf{1}_L, \mathbf{3}_R)$, which we denote by $u_L$ and $u'_L$, respectively. Along with the other two top-like fermions, $t_L'$ and $T_L$, in the $\mathbf{4}$, these fermions are embedded in $\Psi_L$ as
\be
\label{eq:mchm10mass}
\Psi_L 
= \frac1{2} \begin{pmatrix}
  *                                         & \frac{i}2 (u_L+u'_L)   &     *                   &         *                             &     * \\
 -\frac{i}2 (u_L+u'_L)        &  *                                  &     *                    &        *                              &      * \\
 *                                          &  *                                  &    *                     &-\frac{i}2 (u_L-u'_L)      &     -i (t'_{L}+{T}_L) \\
*                                          &    *                                 &\frac{i}2 (u_L-u'_L)   & *                    & (t'_{L}-{T}_L) \\
*                                          & *                                    &i (t'_{L}+{T}_L)  & -(t'_{L}-{T}_L)           &*
\end{pmatrix} ,  
\ee
and similarly for $\Psi_R$. The top Yukawa coupling is
\be
{\cal L}_\Psi = \lambda_m {\rm Tr}(\bar{\Psi}_L \Psi_R) +\lambda_y \Sigma \bar{\Psi}_L \Psi_R \Sigma^T +{\cal L}_{\rm mix}+ {\rm h. c.}\ .
\ee
Including the elementary states, there are five top-like Dirac fermions, $({t}'_L, {T}_L, {t}_L,{u}_L, {u}'_L)$ and $(T_R, t'_R, t_R, u_R , u'_R)$, giving rise to 
\bea
{\rm Det}\left({\cal M}^\dagger{\cal M}\right) &=& \frac1{256} \lambda_1^2 \lambda_2^2\lambda_m^2 (2\lambda_m+ \lambda_y)^2 \, f^{10} \sin^2\frac{2v}{f} \ , \\
\label{eq:mchm10fm}
\frac12 v \frac{\partial}{\partial v}\log {\rm Det}\left({\cal M}^\dagger{\cal M}\right)  &=& 1-\frac43 \frac{v^2}{f^2} \ ,
\eea
which, interestingly, gives the same correction as in MCHM$_5$ in Eq.~(\ref{eq:mchm5fm}). Eq.~(\ref{eq:mchm10fm}) has not been computed previously.
\end{itemize}
\begin{figure}[t]
\includegraphics[scale=0.85]{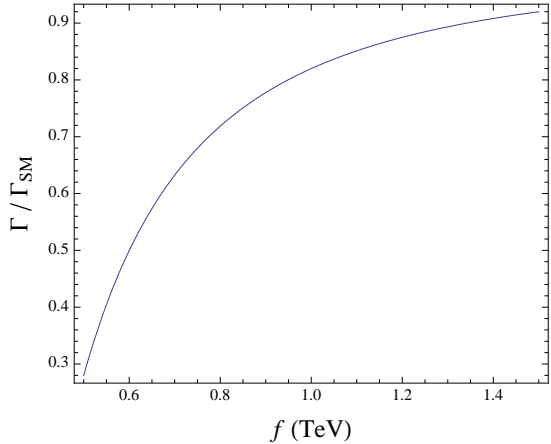} 
\caption{\em The ratio of the Higgs production rates in the gluon fusion channel in the holographic Higgs model over the SM expectation.  With some mild tuning, precision electroweak constraints allow $400$ GeV $\alt f \alt 800$ GeV. \label{fig:fig5} }
\end{figure} 

In the end, we arrive at the result
\be
 {\cal L}_{hgg} =  \frac{g_s^2}{48\pi^2}\frac{h}{v}\left[1-\frac{3}{2}\frac{v^2}{f^2}  \right]G_{\mu\nu}^a G^{a\, \mu\nu} \ .
 \ee
The constraint from the electroweak $S$ parameter prefers $v^2/f^2$ to be in the range of 0.1 -- 0.3 \cite{Contino:2006qr}.\footnote{Such values of $v^2/f^2$ require some mild fine-tuning to achieve. It would be interesting to consider to CFT dual of the warped KK-parity \cite{Agashe:2007jb} to see if the tuning could be reduced.} For example, if $f \sim 700$ GeV, the resulting suppression can be as large as 65\%. In Fig.~\ref{fig:fig5} we plot the ratio of the MCHM predicted rate over the standard model versus $f$.

\section{Summary and discussions}

\label{sect:discuss}

In this work we have provided a general prescription for computing the gluon fusion production of a composite Higgs. Using effective lagrangians the calculations boils down to computing $c_H$, $c_y$, and $c_g$, the coefficients of three dimension-six operators relevant for the Higgs production.  Contributions to $c_H$ come from the nl$\sigma$m as well as integrating out heavy vectors and scalars, which effect is frequently ignored in previous studies. On the other hand, effects of $c_y$ and $c_g$ are easily incorporated by computing the determinant of the fermion mass matrix, without having to solve for the mass eigenvalues and eigenstates. We find that in the end the production rate only depends on the decay constant $f$ and is insensitive to masses of the top partners, which is in sharp contrast with the case of a fundamental Higgs scalar. The effects we considered here are at ${\cal O}(v^2/f^2)$, which {\em a priori} is only at the percent level. However, because all the effects go in the same direction of reducing the production rate, the pile-up effect becomes important and the resulting reduction could be in the range of 10 --  30 \% or greater. It has been claimed that measurements of this production rate at the LHC could have an uncertainty in the order of 20\% \cite{Anastasiou:2005pd}. Our study provides a strong motivation to further reduce the uncertainty in experimental extraction of the Higgs production rate at the LHC.

\begin{figure}[t]
\includegraphics[scale=0.85]{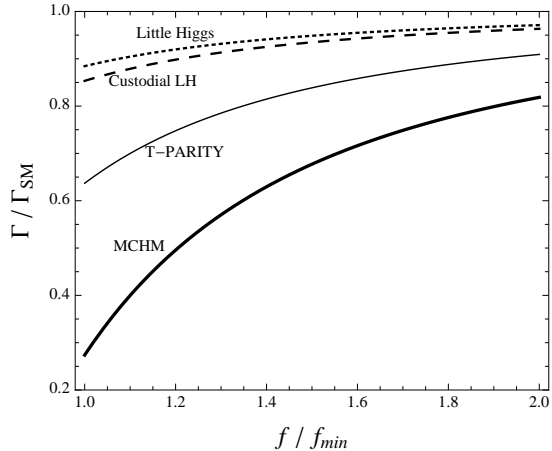} 
\caption{\em The predicted gluon fusion production rates from various composite Higgs models considered in this work. The $f_{min}$, defined as the smallest value allowed by precision electroweak constriants, are: 1.2 TeV (littlest Higgs), 500 GeV (T-parity), 700 GeV (custodial littlest Higgs), and 500 GeV (MCHM). \label{fig:combine} }
\end{figure}

In Fig.~\ref{fig:combine} we show the reduction in the gluon fusion rate for the models considered in this work. We plot the predicted rate over the SM expectation versus the decay constant $f$ in unit of $f_{min}$, which is defined as the smallest value allowed by the precision electroweak constraint. We choose $f_{\min}= 1.2$ TeV, 500 GeV, 700 GeV, and 500 GeV for the littlest Higgs, the T-parity model, the custodial littlest Higgs, and the MCHM, respectively. The reduction is substantial across a wide range of $f$.

Last but not the least, we would like to comment on the effect of higher order QCD corrections, as they are known to be quite large in the particular channel of gluon fusion production \cite{Boughezal:2009fw}. The important observation here is that the strategy of integrating out the heavy colored particles by assuming their masses are much larger than that of the Higgs is validated at the level of NNLO QCD corrections \cite{Marzani:2008az}. Since in models we considered so far the top partners have the same $SU(3)_c$ quantum number as the SM top quark, we speculate that the QCD corrections in the case of integrating out the top partners should be similar to  that from integrating out the SM top quark. As such, the higher order QCD effects should factor out of the effects due to the composite nature of the Higgs. Needless to say, this is an important question deserving further studies, which is nonetheless beyond the scope of the current work.

\begin{acknowledgments}
We acknowledge extensive discussions with R. Rattazzi. One of us (I.~L.) also thanks K. Agashe for clarifications on aspects of the holographic Higgs models. This work is supported in part by the U.S.~Department
of Energy under contracts DE-AC02-06CH11357 and DE-FG02-91ER40684, and by the Swiss National Science Foundation under contract
No. 200021-116372. 

\end{acknowledgments}


\begin{thebibliography}{nn}

\bibitem{:2005ema}
See, for example,   [ALEPH Collaboration and DELPHI Collaboration and L3 Collaboration and OPAL Collaboration and SLD Collaboration and LEP Electroweak Working Group and SLD Electroweak Group and SLD Heavy Flavour Group],
  Phys.\ Rept.\  {\bf 427}, 257 (2006)
  [arXiv:hep-ex/0509008].

\bibitem{Dermisek:2007fi}
  A.~Djouadi,
  Phys.\ Lett.\  B {\bf 435}, 101 (1998)
  [arXiv:hep-ph/9806315]; 
  R.~Dermisek and I.~Low,
  Phys.\ Rev.\  D {\bf 77}, 035012 (2008)
  [arXiv:hep-ph/0701235].


\bibitem{Petriello:2002uu}
  F.~J.~Petriello,
  JHEP {\bf 0205}, 003 (2002)
  [arXiv:hep-ph/0204067].


\bibitem{Boughezal:2010ry}
  R.~Boughezal and F.~Petriello,
  arXiv:1003.2046 [hep-ph]; 
  C.~Anastasiou, R.~Boughezal and E.~Furlan,
  arXiv:1003.4677 [hep-ph].


\bibitem{Boughezal:2009fw}
  R.~Boughezal,
  arXiv:0908.3641 [hep-ph].


\bibitem{Weinberg:1978kz}
  S.~Weinberg,
  Physica A {\bf 96}, 327 (1979).

  
\bibitem{Giudice:2007fh}
  G.~F.~Giudice, C.~Grojean, A.~Pomarol and R.~Rattazzi,
  JHEP {\bf 0706}, 045 (2007)
  [arXiv:hep-ph/0703164].
  
\bibitem{Spira:1995rr}
  M.~Spira, A.~Djouadi, D.~Graudenz and P.~M.~Zerwas,
  Nucl.\ Phys.\  B {\bf 453}, 17 (1995)
  [arXiv:hep-ph/9504378].
  


\bibitem{Kaplan:1983fs}
D.~B.~Kaplan and H.~Georgi,
Phys.\ Lett.\ B {\bf 136}, 183 (1984). 

\bibitem{Kaplan:1983sm}
D.~B.~Kaplan, H.~Georgi and S.~Dimopoulos,
Phys.\ Lett.\ B {\bf 136}, 187 (1984).


\bibitem{Arkani-Hamed:2001nc}
N.~Arkani-Hamed, A.~G.~Cohen and H.~Georgi,
Phys.\ Lett.\ B {\bf 513}, 232 (2001)
[hep-ph/0105239].


\bibitem{Arkani-Hamed:2002qx}
N.~Arkani-Hamed, A.~G.~Cohen, E.~Katz, A.~E.~Nelson, T.~Gregoire and J.~G.~Wacker,
JHEP {\bf 0208}, 021 (2002)
[arXiv:hep-ph/0206020].

\bibitem{Arkani-Hamed:2002qy}
N.~Arkani-Hamed, A.~G.~Cohen, E.~Katz and A.~E.~Nelson,
JHEP {\bf 0207}, 034 (2002)
[arXiv:hep-ph/0206021].

\bibitem{Contino:2003ve}
R.~Contino, Y.~Nomura and A.~Pomarol,
Nucl.\ Phys.\ B {\bf 671}, 148 (2003)
[arXiv:hep-ph/0306259].

\bibitem{Agashe:2004rs}
K.~Agashe, R.~Contino and A.~Pomarol,
 Nucl.\ Phys.\ B {\bf 719}, 165 (2005)
[arXiv:hep-ph/0412089].


\bibitem{Low:2009di}
  I.~Low, R.~Rattazzi and A.~Vichi,
  JHEP {\bf 1004}, 126 (2010)
  [arXiv:0907.5413 [hep-ph]].


\bibitem{Kribs:2007nz}
  G.~D.~Kribs, T.~Plehn, M.~Spannowsky and T.~M.~P.~Tait,
  Phys.\ Rev.\  D {\bf 76}, 075016 (2007)
  [arXiv:0706.3718 [hep-ph]].



\bibitem{Han:2003gf}
  T.~Han, H.~E.~Logan, B.~McElrath and L.~T.~Wang,
  Phys.\ Lett.\  B {\bf 563}, 191 (2003)
  [Erratum-ibid.\  B {\bf 603}, 257 (2004)]
  [arXiv:hep-ph/0302188].
  
  
\bibitem{Chen:2006cs}
  C.~R.~Chen, K.~Tobe and C.~P.~Yuan,
  Phys.\ Lett.\  B {\bf 640}, 263 (2006)
  [arXiv:hep-ph/0602211].


\bibitem{Djouadi:2007fm}
  A.~Djouadi and G.~Moreau,
  Phys.\ Lett.\  B {\bf 660}, 67 (2008)
  [arXiv:0707.3800 [hep-ph]].


\bibitem{Falkowski:2007hz}
  A.~Falkowski,
  Phys.\ Rev.\  D {\bf 77}, 055018 (2008)
  [arXiv:0711.0828 [hep-ph]].

\bibitem{Maru:2010ic}
  N.~Maru and N.~Okada,
  arXiv:1002.1835 [hep-ph].


\bibitem{Espinosa:2010vn}
  J.~R.~Espinosa, C.~Grojean and M.~Muhlleitner,
  JHEP {\bf 1005}, 065 (2010)
  [arXiv:1003.3251 [hep-ph]].

\bibitem{Azatov:2010pf}
  A.~Azatov, M.~Toharia and L.~Zhu,
  Phys.\ Rev.\  D {\bf 82}, 056004 (2010)
  [arXiv:1006.5939 [hep-ph]].


\bibitem{Casagrande:2010si}
  S.~Casagrande, F.~Goertz, U.~Haisch, M.~Neubert and T.~Pfoh,
  arXiv:1005.4315 [hep-ph].
  
  
    
\bibitem{Gunion:2002zf}
  J.~F.~Gunion and H.~E.~Haber,
  Phys.\ Rev.\  D {\bf 67}, 075019 (2003)
  [arXiv:hep-ph/0207010].

  
\bibitem{Coleman:sm}
S.~R.~Coleman, J.~Wess and B.~Zumino,
Phys.\ Rev.\  {\bf 177} (1969) 2239.

\bibitem{Callan:sn}
C.~G.~.~Callan, S.~R.~Coleman, J.~Wess and B.~Zumino,
Phys.\ Rev.\  {\bf 177} (1969) 2247.


\bibitem{Barbieri:2004qk}
  R.~Barbieri, A.~Pomarol, R.~Rattazzi and A.~Strumia,
  Nucl.\ Phys.\  B {\bf 703}, 127 (2004)
  [arXiv:hep-ph/0405040].
  
\bibitem{Ellis:1975ap}  
J.~R.~Ellis, M.~K.~Gaillard and D.~V.~Nanopoulos,
Nucl.\ Phys.\ B {\bf 106}, 292 (1976).

\bibitem{Shifman:1979eb}
  M.~A.~Shifman, A.~I.~Vainshtein, M.~B.~Voloshin and V.~I.~Zakharov,
  Sov.\ J.\ Nucl.\ Phys.\  {\bf 30}, 711 (1979)
  [Yad.\ Fiz.\  {\bf 30}, 1368 (1979)].


\bibitem{Cheng:2003ju}
H.~C.~Cheng and I.~Low,
JHEP {\bf 0309}, 051 (2003)
[arXiv:hep-ph/0308199];
H.~C.~Cheng and I.~Low,
JHEP {\bf 0408}, 061 (2004)
  [arXiv:hep-ph/0405243];
I.~Low,
JHEP {\bf 0410}, 067 (2004)
[arXiv:hep-ph/0409025].

\bibitem{Pappadopulo:2010jx}
  D.~Pappadopulo and A.~Vichi,
  arXiv:1007.4807 [hep-ph].





  
  





\bibitem{Chang:2003zn}
  S.~Chang,
  JHEP {\bf 0312}, 057 (2003)
  [arXiv:hep-ph/0306034].


\bibitem{Contino:2006qr}
  R.~Contino, L.~Da Rold and A.~Pomarol,
  Phys.\ Rev.\  D {\bf 75}, 055014 (2007)
  [arXiv:hep-ph/0612048].
  
\bibitem{Agashe:2006at}
  K.~Agashe, R.~Contino, L.~Da Rold and A.~Pomarol,
  Phys.\ Lett.\  B {\bf 641}, 62 (2006)
  [arXiv:hep-ph/0605341].


\bibitem{Csaki:2002qg}
  C.~Csaki, J.~Hubisz, G.~D.~Kribs, P.~Meade and J.~Terning,
  Phys.\ Rev.\  D {\bf 67}, 115002 (2003)
  [arXiv:hep-ph/0211124].

\bibitem{Csaki:2003si}
  C.~Csaki, J.~Hubisz, G.~D.~Kribs, P.~Meade and J.~Terning,
  Phys.\ Rev.\  D {\bf 68}, 035009 (2003)
  [arXiv:hep-ph/0303236].

\bibitem{Schmaltz:2008vd}
  M.~Schmaltz and J.~Thaler,
  JHEP {\bf 0903}, 137 (2009)
  [arXiv:0812.2477 [hep-ph]].
  
\bibitem{Surujon:2010ed}
  Z.~Surujon and P.~Uttayarat,
  arXiv:1003.4779 [hep-ph].

  
 


\bibitem{Contino:2006nn}
  R.~Contino, T.~Kramer, M.~Son and R.~Sundrum,
  JHEP {\bf 0705}, 074 (2007)
  [arXiv:hep-ph/0612180].

\bibitem{Maldacena:1997re}
  J.~M.~Maldacena,
  Adv.\ Theor.\ Math.\ Phys.\  {\bf 2}, 231 (1998)
  [arXiv:hep-th/9711200].


\bibitem{Barbieri:2007bh}
  R.~Barbieri, B.~Bellazzini, V.~S.~Rychkov and A.~Varagnolo,
  Phys.\ Rev.\  D {\bf 76}, 115008 (2007)
  [arXiv:0706.0432 [hep-ph]].

\bibitem{Agashe:2007jb}
  K.~Agashe, A.~Falkowski, I.~Low and G.~Servant,
  JHEP {\bf 0804}, 027 (2008)
  [arXiv:0712.2455 [hep-ph]].

\bibitem{Anastasiou:2005pd}
  C.~Anastasiou, K.~Melnikov and F.~Petriello,
  Phys.\ Rev.\  D {\bf 72}, 097302 (2005)
  [arXiv:hep-ph/0509014].





  
\bibitem{Marzani:2008az}
  S.~Marzani, R.~D.~Ball, V.~Del Duca, S.~Forte and A.~Vicini,
  Nucl.\ Phys.\  B {\bf 800}, 127 (2008)
  [arXiv:0801.2544 [hep-ph]];
  R.~V.~Harlander, H.~Mantler, S.~Marzani and K.~J.~Ozeren,
  Eur.\ Phys.\ J.\  C {\bf 66}, 359 (2010)
  [arXiv:0912.2104 [hep-ph]].


  




\end{thebibliography}
\end{document}